 \documentclass{vgtc}                          

\graphicspath{{figures/}{pictures/}{images/}{./}} 

\usepackage{times}                     

\usepackage{tabu}                      
\usepackage{booktabs}                  
\usepackage{lipsum}                    
\usepackage{mwe}                       

\usepackage{mathptmx}                  

\usepackage{graphicx}
\usepackage{makecell}

\onlineid{0}

\vgtccategory{Research}

\vgtcinsertpkg

\title{How Early Can You Tell? Early Eye Gaze Dynamics and Cybersickness Progression in Virtual Reality}

\author{Nevzat Umut Demirseren\thanks{e-mail: ndemirseren@txstate.edu}\\ %
        \scriptsize Texas State University %
\and Isayas Berhe Adhanom\thanks{e-mail: isayas@txstate.edu}\\ %
     \scriptsize Texas State University}

\abstract{
    Cybersickness remains one of the primary barriers to prolonged and comfortable virtual reality (VR) use, yet the temporal development of cybersickness remains poorly understood because its onset and progression vary substantially across individuals. Eye tracking provides a continuous and unobtrusive measure of user behavior, making it well suited for investigating how users respond throughout VR exposure. However, existing work has mainly focused on aggregate gaze measures or post-exposure assessments, leaving the relationship between early eye-movement dynamics and discomfort progression underexplored. In this paper, we investigate whether early eye-movement dynamics characterize the subsequent progression of cybersickness. Using trajectories of fixation duration, fixation dispersion, saccade amplitude, saccade velocity, and gaze eccentricity computed over the first one to five minutes of VR exposure, we evaluate their association with subsequent discomfort progression and compare early gaze dynamics between participants with low and high post-exposure cybersickness severity. Our results show that early fixation dispersion is consistently associated with subsequent discomfort progression, whereas no significant differences are observed between cybersickness severity groups. These findings suggest that early fixation dispersion can characterize the progression of discomfort rather than the final cybersickness severity. Overall, this work provides new insight into the temporal relationship between early eye-movement behavior and cybersickness progression, contributing to the understanding of gaze dynamics during immersive experiences.
} 


\begin{document}


\firstsection{Introduction}

\maketitle


Virtual reality (VR) technologies have advanced rapidly in recent years, enabling immersive applications across entertainment, education, healthcare, training, and scientific visualization. As VR becomes increasingly integrated into these domains, providing users with experiences that remain comfortable over extended periods has become increasingly important. However, prolonged immersion can introduce challenges that negatively affect user experience and limit the practical adoption of VR systems.

Among these challenges, cybersickness remains one of the primary barriers to the widespread adoption of VR \cite{stanney2020identifying,tian2022review}. During VR exposure, users may develop symptoms such as nausea, dizziness, and disorientation, which reduce comfort, impair performance, and limit the duration of immersive experiences \cite{kennedy1993simulator}. Although these symptoms are commonly experienced during VR exposure, their severity and progression vary considerably across individuals, making it difficult to anticipate when discomfort will arise or how it will evolve throughout an experience. Consequently, estimating, predicting, and mitigating cybersickness remain active areas of research. This challenge has motivated increasing interest in adaptive VR systems capable of continuously monitoring user state and responding to behavioral changes as they occur, with the goal of improving user comfort and supporting longer and more effective immersive experiences.

To continuously monitor user state, modern VR systems increasingly leverage integrated sensing technologies, including eye trackers, head and hand tracking, and environmental context \cite{adhanom2023eye, bozkir2026eye, Clay2019-ke, plopski2022eye}. These sensing capabilities enable immersive systems to continuously interpret cognitive, perceptual, and behavioral user states, creating opportunities for adaptive and personalized experiences \cite{Jain2023, Li2025AdaptingTT}. Among the available sensing modalities, eye tracking has emerged as a particularly promising approach because it provides a rich measure of user behavior without disrupting the immersive experience. Eye movements have been associated with a wide range of user states, including visual attention, cognitive workload, fatigue, task engagement, and cybersickness, making gaze an attractive modality for continuous behavioral sensing in VR \cite{fan2023eye, jeong2022eyes, liu2022assessing, ozkan2023relationship, walter2021cognitive}. Existing work has demonstrated that gaze behavior reflects changes in user state using eye-movement metrics and data-driven learning approaches \cite{chang_predicting_2021, islam_cybersickness_2021, kundu2023litevr}.

Despite these advances, most studies characterize gaze using aggregate session-level measures or evaluate behavior after substantial changes in user state have already occurred, providing limited insight into how eye-movement behavior evolves during the earliest stages of VR exposure. Eye movements continuously adapt as users perceive, explore, and interact with virtual environments, reflecting ongoing changes in visual processing and sensorimotor behavior. During the initial minutes of VR exposure, users begin adapting to the visual and vestibular demands imposed by the virtual environment while the earliest symptoms of cybersickness may also begin to emerge \cite{laviola2000discussion, stanney2020identifying}. Consequently, changes in gaze behavior during this period may capture the onset of behavioral adaptations that precede more pronounced discomfort later in the experience. Examining gaze during early exposure may therefore provide insight into how users adapt to immersive environments before the effects of prolonged exposure, accumulated fatigue, and sustained discomfort become more pronounced. Such an understanding could improve our knowledge of cybersickness development while informing gaze-based user-state assessment for adaptive VR systems. Consequently, it remains unclear whether early gaze dynamics are associated with the subsequent progression of discomfort or differ between users who ultimately experience different levels of cybersickness.

In this paper, we investigate the relationship between early eye-movement dynamics and cybersickness through two complementary analyses. We first examine whether gaze trajectories computed over progressively longer early exposure periods (1--5 minutes) are associated with subsequent trajectories of self-reported discomfort throughout the remainder of the VR experience. We then compare early gaze trajectories between participants with low and high cybersickness severity, defined using post-completion sickness scores. Across these analyses, we characterize the temporal evolution of fixation duration, fixation dispersion, saccade amplitude, saccade velocity, and gaze eccentricity during the early stages of VR exposure. Our results demonstrate that early changes in fixation dispersion are associated with subsequent discomfort progression, while providing no evidence that early gaze trajectories differ according to overall post-exposure cybersickness severity. Overall, these findings provide new insight into the temporal relationship between early oculomotor behavior, discomfort progression, and post-exposure cybersickness severity, contributing to a better understanding of gaze behavior during the earliest stages of VR exposure and informing future gaze-based adaptive VR systems.

\section{Background}

\subsection{Cybersickness in Virtual Reality}

Cybersickness is commonly described as a form of visually induced motion sickness and remains one of the primary barriers limiting prolonged and comfortable VR use. It is characterized by symptoms of nausea, oculomotor discomfort, and disorientation, which can reduce user comfort, impair task performance, and shorten the duration of immersive experiences \cite{kennedy1993simulator, laviola2000discussion, rebenitsch2016review}. To mitigate these effects, prior work has proposed a range of techniques, such as dynamic field-of-view (FOV) restriction, peripheral vignetting, visual rest frames, and adaptive locomotion strategies that reduce visual stimuli associated with cybersickness while attempting to preserve immersion \cite{fernandes2016combating, rebenitsch2016review}. Although these approaches can improve user comfort, they often introduce trade-offs with visual fidelity and immersion, motivating continued research into adaptive behavioral approaches that respond to changes in user state during VR exposure.

\subsection{Causes of Cybersickness}

Although no single theory fully explains cybersickness \cite{stanney2020identifying, tian2022review}, several established frameworks have been proposed to explain its underlying mechanisms.

\subsubsection{Sensory Conflict Theory}

Sensory Conflict Theory is one of the most widely accepted explanation of cybersickness \cite{oman_motion_1990, reason1975motion}. It proposes that discomfort arises when visual information conflicts with vestibular and proprioceptive cues. During VR exposure, users may perceive visually induced self-motion without corresponding physical motion, producing sensory conflict that contributes to cybersickness.

\subsubsection{Postural Instability Theory}

Postural Instability Theory attributes cybersickness to prolonged difficulty maintaining stable postural control \cite{riccio1991ecological, stoffregen1998postural}. According to this perspective, instability precedes the onset of cybersickness, and symptoms persist until stable postural control is reestablished.

\subsubsection{Eye Movement Theory}

Eye Movement Theory suggests that cybersickness arises from excessive oculomotor strain induced by visual motion \cite{ebenholtz_motion_1992, ebenholtz_possible_1994}. Sustained demands on gaze stabilization and eye-movement control during VR exposure may therefore alter fixation and saccadic behavior, motivating the investigation of eye-movement dynamics as behavioral indicators of cybersickness.

\subsection{Cybersickness Assessment}

\subsubsection{Subjective Measures}

Cybersickness is most commonly assessed using subjective self-report measures. The Simulator Sickness Questionnaire (SSQ) provides a standardized post-exposure evaluation of nausea, oculomotor discomfort, and disorientation symptoms \cite{kennedy1993simulator}. Complementary measures, including the Fast Motion Sickness (FMS) scale \cite{keshavarz2011validating} and repeated in-experience discomfort ratings \cite{adhanom2020effect, fernandes2016combating}, enable repeated assessments that capture the progression of symptoms throughout VR exposure. Despite their widespread use, these measures rely on user self-report.

\subsubsection{Objective Measures}

To complement subjective assessments, researchers have explored objective indicators of cybersickness derived from physiological and behavioral signals, including electrodermal activity (EDA), heart rate, blink activity, postural sway, and VR tracking data \cite{arcioni2019postural, gervsak2020effect, lopes_eye_2020, palmisano2018predicting}. These measures provide continuous assessments of user state and have demonstrated relationships with cybersickness throughout VR exposure. Among these objective measures, eye movements have emerged as a particularly promising behavioral signal because they directly reflect changes in visual exploration and oculomotor behavior, motivating their investigation as indicators of cybersickness.

\subsection{Eye Movements and Cybersickness}

Eye movements provide a direct window into visual processing and oculomotor control, making them well suited for investigating behavioral changes associated with cybersickness. Previous work has shown that fixation characteristics, saccade behavior, blink activity, pupil responses, and other gaze-derived measures are associated with discomfort during VR exposure, demonstrating that eye movements contain meaningful information related to cybersickness \cite{nam_eye_2022, ozkan2023relationship}. Beyond cybersickness, eye movements have also been widely used to characterize user state, including visual attention, cognitive workload, fatigue, and task engagement, highlighting their value as continuous behavioral indicators in immersive environments \cite{foucher2025eye, wang2025microsaccades}.

Building on these observations, several studies have leveraged eye tracking to model or predict cybersickness, either using gaze features alone or in combination with other behavioral and physiological signals \cite{chang_predicting_2021, dennison2016use, islam_cybersickness_2021, kundu2023litevr, lopes_eye_2020}. While these approaches demonstrate that eye movements are informative indicators of discomfort, they have primarily relied on aggregate session-level measures, condition-based comparisons, or predictive modeling frameworks. Consequently, relatively little attention has been given to characterizing how eye-movement behavior evolves during the earliest stages of VR exposure or whether these early behavioral dynamics are associated with subsequent discomfort progression and overall cybersickness severity. Addressing this gap may improve our understanding of the temporal development of cybersickness while informing gaze-based user-state assessment for future adaptive VR systems.

\section{Methodology}

\subsection{Dataset}

The present work builds upon a previously published study \cite{adhanom2022vr}. The original study was approved by the Institutional Review Board (IRB), and all participants provided informed consent prior to participation. The following sections summarize the experimental protocol, measurements, and data processing procedures relevant to the current analysis, while complete details of the original study are available in the original publication.

\subsubsection{Experiment Design}

The original study consisted of a prolonged VR locomotion task where participants navigated a virtual environment by following visual waypoints using controller-based locomotion and head-steered navigation. Each session lasted up to 20 minutes, allowing repeated measurements of discomfort throughout VR exposure. The dataset included two sub-experiments conducted in realistic and abstract virtual environments. Consistent with our research objective, we restricted the present analysis to the  realistic environment session, as it provides naturalistic visual exploration while avoiding potential adaptation effects associated with repeated VR exposure, thereby offering a clearer characterization of discomfort progression during a single continuous session.

\begin{figure}[htbp]
  \centering
  \includegraphics[width=1.00\linewidth]{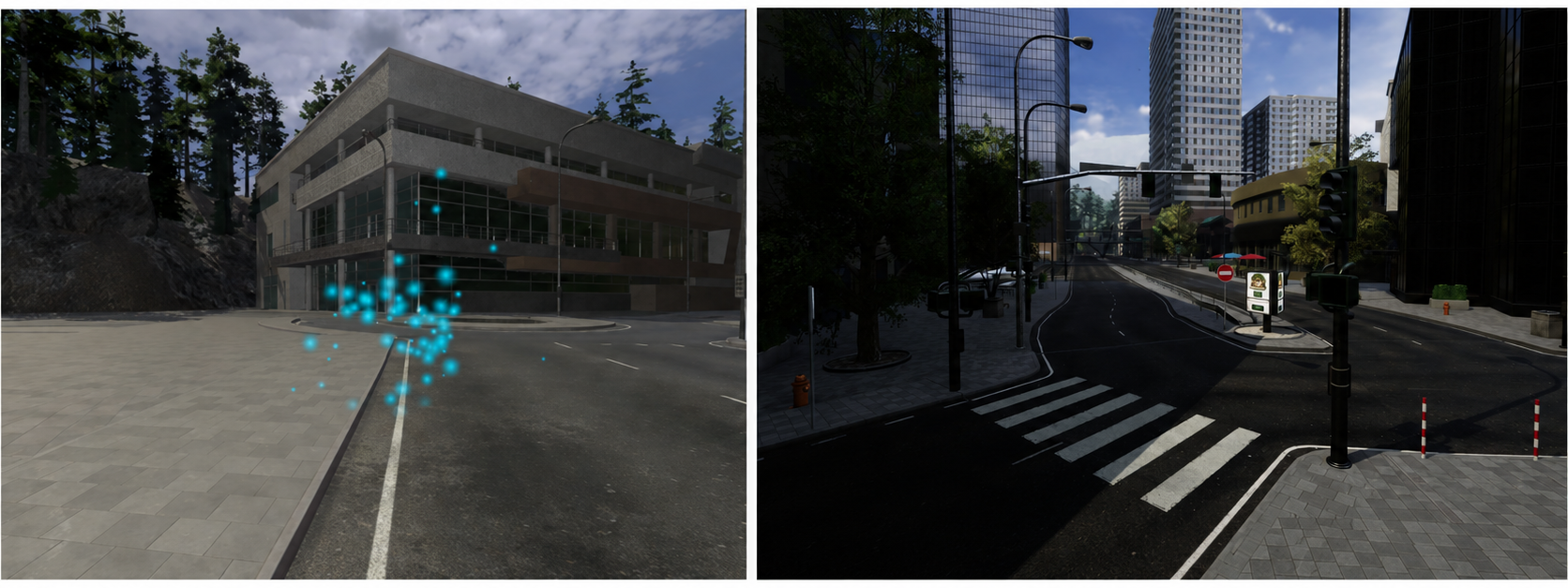}
  \caption{Representative screenshots from the experiment's virtual environment used for the study.}
  \label{fig:environment}
\end{figure}

\subsubsection{Procedure}

Participants first completed informed consent, a demographic questionnaire, and a baseline SSQ before receiving instructions on the VR navigation task. Afterwards, the participants navigated a virtual environment for approximately 20 minutes while remaining stationary in the physical space, using an Xbox 360 controller for locomotion and head-steered navigation for directional control. The environment permitted natural, unconstrained visual exploration without requiring gaze-contingent tasks. During the session, participants periodically reported their discomfort on a 0--10 visual rating scale, enabling continuous assessment of symptom progression. Following the navigation task, participants completed a post-exposure SSQ to assess overall cybersickness severity.

\subsubsection{Participants}

The original study enrolled 22 participants (11 male, 11 female; mean age = 23.45 years, SD = 3.9). The present analysis included data from the 19 participants (11 male, 8 female; mean age = 22.8 years, SD = 3.6) who completed the study.

\subsubsection{Measures}

Cybersickness was assessed using complementary subjective measures. Participants completed the SSQ before and after VR exposure to quantify overall cybersickness severity, while discomfort was assessed throughout the session using a 0--10 rating scale administered at one-minute intervals. The repeated discomfort ratings provided a time-resolved measure of symptom progression during the VR experience, complementing the overall pre- and post-exposure SSQ assessments.

\subsection{Study Design}


\subsubsection{Eye Movement Processing Pipeline}

Raw gaze samples were acquired at 120 Hz using the integrated eye tracker of the HTC Vive Pro Eye head-mounted display (HMD). Gaze direction from the left eye was used throughout the analysis to provide a consistent monocular signal for event-level processing \cite{hooge2019gaze}.
Samples flagged as invalid were removed, gaps shorter than 75 ms were linearly interpolated to preserve temporal continuity, and the resulting gaze signal was smoothed using a five-sample moving average filter to reduce high-frequency measurement noise.

Angular displacement was computed between consecutive 3D gaze direction vectors expressed in the head-relative (eye-in-head) coordinate system, and angular velocity was obtained by dividing this displacement by the corresponding time interval between samples.
Eye-movement events were subsequently identified using the velocity-threshold (I-VT) algorithm \cite{salvucci2000identifying}, following established parameter selection guidelines \cite{holmqvist2011eye, olsen2012identifying}.
Samples with angular velocity exceeding $35^\circ$/s were classified as saccades, whereas lower-velocity samples were considered candidate fixations. Consecutive fixation segments separated by less than 75 ms and an angular distance below $0.5^\circ$ were merged to account for brief interruptions in otherwise continuous fixations.

To improve the robustness of event detection, duration-based filtering was applied to all detected events. Fixations were required to have durations between 60 ms and 2000 ms, while saccades were constrained to durations between 10 ms and 150 ms with a minimum amplitude of $1^\circ$.

From the resulting event stream, we computed event-level eye-movement metrics that summarize physiologically meaningful aspects of gaze behavior. By characterizing fixation and saccade events rather than individual gaze samples, these metrics provide behaviorally interpretable measures of how users allocate attention, maintain gaze stability, and explore the virtual environment. Accordingly, we selected complementary fixation- and saccade-based metrics that capture oculomotor processes associated with visual behavior and cybersickness \cite{foucher2025eye,ozkan2023relationship}.

Five eye-movement metrics were extracted from the detected events. For fixation events, we computed mean fixation duration, which reflects the time spent processing visual information, and fixation dispersion, which characterizes the spatial stability and spread of gaze within a fixation. For saccades, we computed mean saccade amplitude and mean saccade velocity, capturing the extent and dynamics of visual exploration between fixation locations. We additionally computed gaze eccentricity as the angular distance between the gaze direction and the center of the visual field, representing how far visual attention is directed from the forward viewing direction. Metric definitions and computation procedures follow established eye-tracking methodologies \cite{holmqvist2011eye}.

\subsubsection{Discomfort Trajectory Assessment}

To investigate whether early gaze behavior is associated with the progression of cybersickness, we analyzed early exposure periods spanning the first 1--5 minutes of VR exposure. Because cybersickness develops over the course of immersion, repeated discomfort measurements provide a more informative characterization of symptom evolution than post-exposure assessments alone \cite{garrido2022focusing, ryu2026predicting}. Focusing on the initial stages of VR exposure enables us to examine whether early eye-movement dynamics are associated with the evolution of discomfort before prolonged exposure may further influence gaze behavior.

Eye tracking data were partitioned into consecutive 30-second windows, and event-level gaze metrics were averaged within each window to construct early gaze trajectories. A 30-second window was selected based on a sensitivity analysis comparing multiple temporal granularities, as it provided the most stable characterization of gaze trajectories while preserving temporal variation during early VR exposure. Shorter windows produced noisier trajectory estimates due to fewer eye-movement events, whereas longer windows excessively smoothed temporal changes. To determine how much early gaze behavior is required before meaningful associations with subsequent discomfort progression emerge, trajectories were computed using progressively increasing early exposure durations (0--1, 0--2, 0--3, 0--4, and 0--5 minutes). For each early exposure duration, the trajectory of each gaze metric was quantified as the linear slope across the corresponding windows. Discomfort trajectory was computed from the remaining discomfort ratings collected after the corresponding early exposure period.

\subsubsection{Cybersickness Severity Grouping}

To examine whether early gaze behavior differed according to overall cybersickness severity, participants were grouped using the difference between their post-exposure and baseline SSQ scores. As the primary analysis, participants were divided using a relative SSQ threshold of 40, consistent with guidelines for moderate-to-severe cybersickness grouping \cite{kelly2026interpreting}. Participants with relative SSQ scores below 40 were assigned to the low-severity group, whereas those with scores of 40 or greater were assigned to the high-severity group. Because no universally accepted binary SSQ threshold exists, we repeated the analysis using a sample-based median split of the relative SSQ scores (median = 44.88) as a sensitivity analysis to assess whether the findings were robust to the choice of grouping strategy. For both grouping approaches, comparisons were performed for gaze trajectories computed over each progressively increasing early exposure interval (0--1, 0--2, 0--3, 0--4, and 0--5 minutes).

\subsubsection{Statistical Analysis}

Two complementary analyses were conducted to evaluate the relationship between early gaze dynamics and cybersickness from two perspectives. The first analysis examined whether individual differences in early eye-movement trajectories were associated with the subsequent progression of discomfort throughout the remainder of the VR session. The second analysis investigated whether early gaze behavior differed between participants who ultimately experienced different overall cybersickness severity, thereby characterizing how early eye-movement dynamics varied across participants with different symptom outcomes. Across all statistical analyses, $p$-values were adjusted using the Holm--Bonferroni procedure \cite{holm1979simple} to control the family-wise error rate.

For the first analysis, linear regression models were used to examine whether trajectories of eye-movement features during the early stages of VR exposure were associated with subsequent discomfort progression. Separate models were fit for each eye-movement metric and early exposure duration using both the future discomfort trajectory and a normalized progression index as outcome variables. The normalized progression index was computed as the ratio of the future discomfort slope to the overall discomfort slope.

The linear models are formulated as:

\[
\mathrm{Outcome}_i =
\beta_0 +
\beta_1\,\mathrm{EarlySlope}_i +
\varepsilon_i
\]

where $\mathrm{Outcome}$ denotes either the participant-specific future discomfort trajectory or the normalized progression index. $\mathrm{EarlySlope}$ represents the linear slope of an eye-movement metric computed across the corresponding early exposure duration, while the future discomfort trajectory is defined as the linear slope of the discomfort ratings collected after the early exposure period. $\beta_0$ is the intercept, $\beta_1$ is the regression coefficient, and $\varepsilon_i$ is the residual error.

The second analysis focused on overall cybersickness severity by comparing early gaze dynamics between participants in the low- and high-severity cybersickness groups. For each early exposure duration, the trajectories of each eye-movement metric were compared between the two groups using the Mann--Whitney $U$ test \cite{mann1947test}. Unlike the regression analysis, which evaluated whether early gaze trajectories were associated with subsequent discomfort progression, this analysis examined whether participants with different overall levels of post-exposure cybersickness exhibited distinct eye-movement dynamics during the initial stages of VR exposure.

\section{Results}

\subsection{Discomfort Progression}

For the future discomfort trajectory, fixation dispersion was significantly associated with subsequent discomfort progression when early gaze trajectories were computed over the first two minutes ($\beta=-1.52$, $p_{\mathrm{Holm}}=.0085$, $R^2=.45$), the first three minutes ($\beta=-3.99$, $p_{\mathrm{Holm}}=.0166$, $R^2=.41$), and the first four minutes ($\beta=-3.37$, $p_{\mathrm{Holm}}=.0312$, $R^2=.36$). The association was no longer significant when the first five minutes of early gaze data were included ($p_{\mathrm{Holm}}=.151$).

For the normalized progression index, fixation dispersion was significantly associated with discomfort progression when early gaze trajectories were computed over the first four minutes ($\beta=-18.24$, $p_{\mathrm{Holm}}=.0013$, $R^2=.55$) and the first five minutes ($\beta=-27.06$, $p_{\mathrm{Holm}}=.0254$, $R^2=.38$). Although fixation dispersion approached significance after the first three minutes ($\beta=-11.41$, $p_{\mathrm{Holm}}=.051$, $R^2=.33$), it did not remain significant following Holm--Bonferroni correction.

No other eye-movement metric exhibited statistically significant associations after Holm--Bonferroni correction. Prior to correction, significant results were observed for saccade amplitude (\(\beta=-0.13\), \(p=.035\)) and saccade velocity (\(\beta=-0.013\), \(p=.019\)) with the future discomfort trajectory using the first three minutes of early gaze data. For the normalized progression index, fixation dispersion also exhibited a significant association before correction after the first three minutes (\(\beta=-11.41\), \(p=.010\)).

\begin{figure*}[t]
    \centering

    \begin{minipage}[t]{0.495\textwidth}
        \centering
        \includegraphics[width=\linewidth]{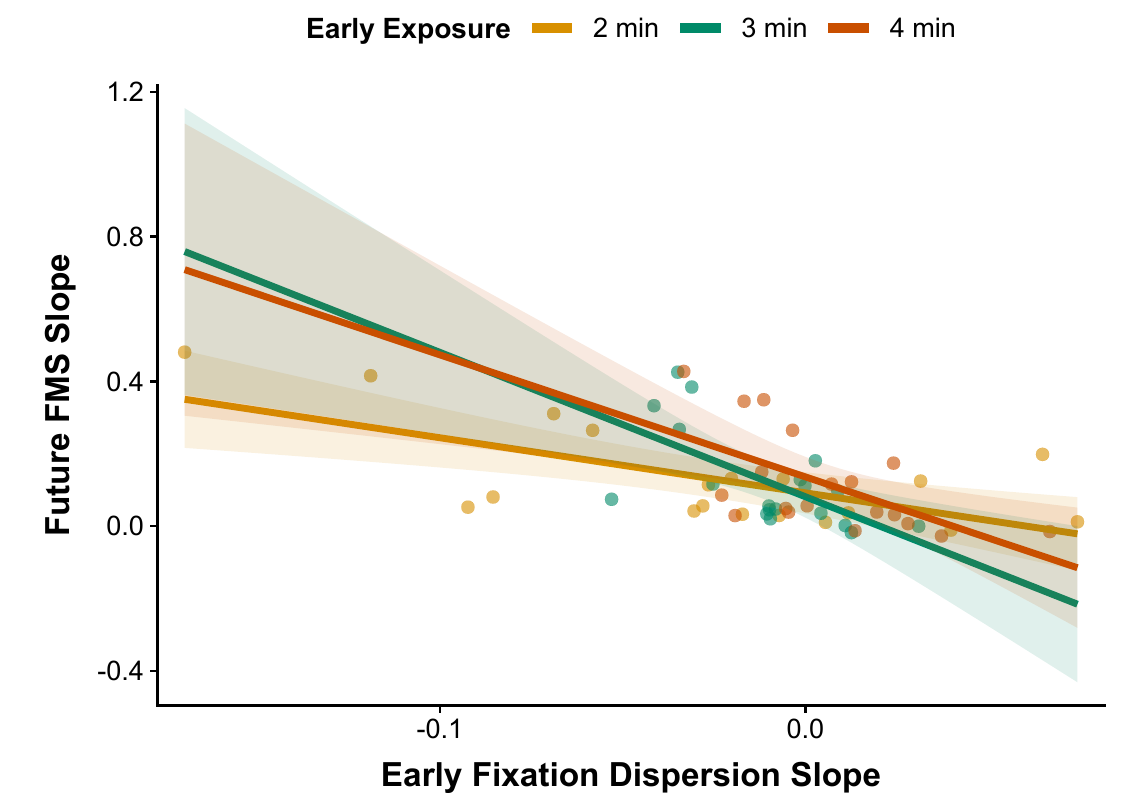}
        \small (a) Future discomfort trajectory
    \end{minipage}
    \hfill
    \begin{minipage}[t]{0.495\textwidth}
        \centering
        \includegraphics[width=\linewidth]{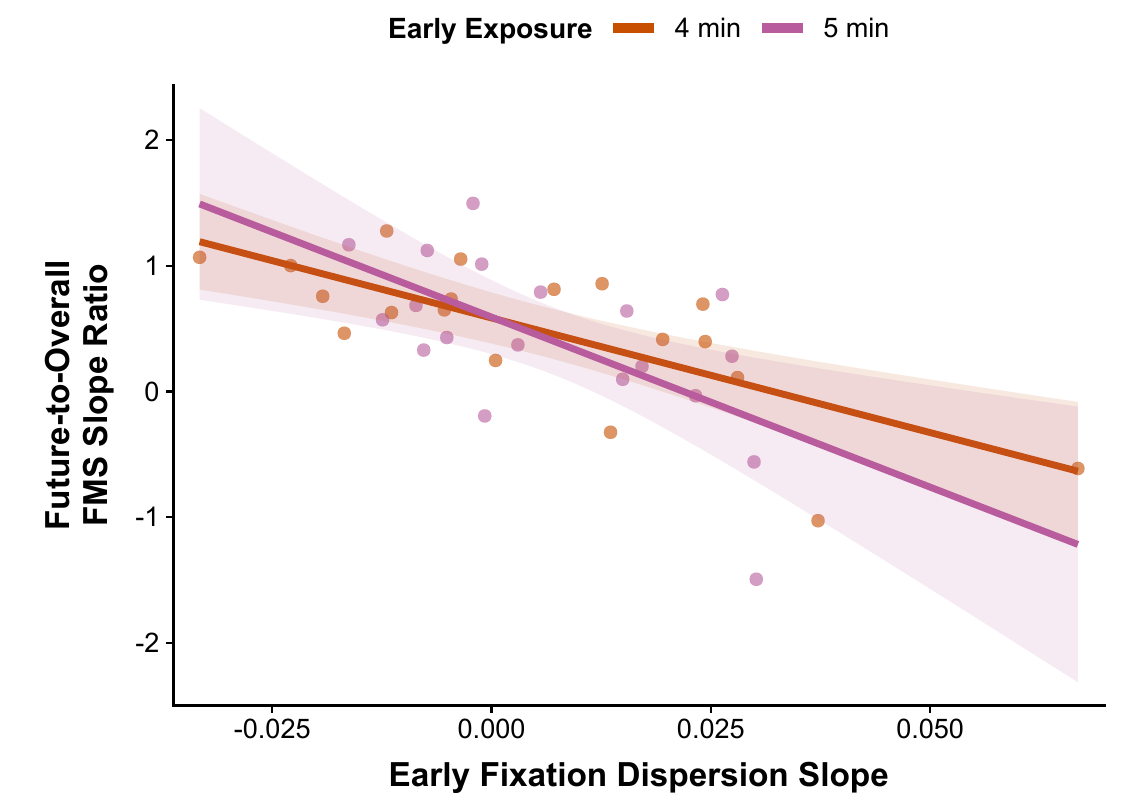}
        \small (b) Normalized progression index
    \end{minipage}

    \caption{
        Participant-level associations between early fixation dispersion
        trajectories and subsequent discomfort progression. Points represent
        individual participants, solid lines show fitted linear regression
        models for the significant early-exposure durations, and shaded regions
        denote 95\% confidence intervals. Colors distinguish the early-exposure
        durations included in each model.
    }
    \label{fig:fixation_dispersion_scatter}
\end{figure*}

Figure~\ref{fig:fixation_dispersion_scatter}a illustrates the relationships between early fixation dispersion trajectories and the subsequent discomfort trajectory and Figure~\ref{fig:fixation_dispersion_scatter}b shows the corresponding relationships with the normalized progression index.

\subsection{Cybersickness Severity Group Comparisons}

Using the SSQ threshold of 40, the low- and high-severity groups comprised 8 and 11 participants, respectively (low: median = 13.09, IQR = 10.29--19.63; high: median = 56.10, IQR = 48.62--69.19). Using the median split, the corresponding group sizes were 9 and 10 participants (low: median = 14.96, IQR = 11.22--22.44; high: median = 56.10, IQR = 49.55--75.73). The two grouping strategies differed in the classification of one participant.

Across both grouping strategies, no significant differences in early gaze trajectories were observed between the low- and high-severity cybersickness groups for any eye-movement metric or early exposure duration. Using the first threshold, the smallest uncorrected \(p\)-value was observed for fixation dispersion during the first two minutes of exposure (\(U=22\), \(p=.076\), \(p_{\mathrm{Holm}}=.379\)); all remaining comparisons yielded uncorrected \(p>.12\). Repeating the analysis using the median split produced the same overall conclusion, with no significant between-group differences for any eye-movement metric or early exposure duration.

\section{Discussion}

This study investigated whether trajectories of event-level eye-movement metrics during the initial stages of VR exposure are associated with discomfort progression and post-exposure cybersickness severity. Overall, the results indicate that early trajectories of fixation dispersion were consistently associated with subsequent discomfort progression across multiple early exposure durations, whereas fixation duration, saccade amplitude, saccade velocity, and gaze eccentricity showed no consistent associations after correction for multiple comparisons. Furthermore, no statistically significant differences in early gaze trajectories were observed between participants grouped according to post-exposure cybersickness severity.
These findings suggest that early gaze behavior is more closely related to discomfort progression than to final symptom severity. 

To better understand the observed relationship between fixation dispersion and subsequent discomfort progression, it is useful to consider the behavioral characteristics captured by this metric.
Fixation dispersion reflects the spatial stability of gaze during individual fixations. Increasing fixation dispersion indicates that gaze becomes less spatially constrained over time, whereas decreasing fixation dispersion reflects progressively more stable and localized fixations. Rather than reflecting a deliberate viewing strategy, fixation dispersion trajectories may capture differences in how users' oculomotor behavior adapts during the initial stages of VR exposure as sensory conflict accumulates. The observed  association with subsequent discomfort progression suggests that fixation dispersion captures dynamic behavioral changes accompanying the development of cybersickness, rather than a fixed characteristic of an individual's gaze behavior.

One possible account is that the observed reduction in fixation dispersion reflects a compensatory oculomotor response to accumulating sensory conflict. Eye Movement Theory holds that visually induced motion sickness arises in part from the demands visual motion places on gaze stabilization~\cite{ebenholtz_motion_1992, ebenholtz_possible_1994}; participants whose dispersion narrowed early may have been increasingly constraining gaze to stabilize the retinal image as visual--vestibular conflict accumulated, and these participants subsequently exhibited steeper discomfort increases. A less specific alternative is that the trajectories simply index individual differences in how users adapt to the visual and vestibular demands of early immersion, consistent with prior reports that fixation behavior relates to cybersickness~\cite{ozkan2023relationship} and that gaze patterns evolve as symptoms develop~\cite{nam_eye_2022}.  

Taken together, these findings suggest that eye-movement behavior during the first few minutes of VR exposure may provide information about the temporal evolution of cybersickness rather than its ultimate severity. In particular, early fixation dispersion trajectories appear to capture differences in discomfort progression during immersive VR exposure, even when users ultimately report comparable post-exposure symptom levels. These findings highlight the potential value of early gaze-based behavioral measures for informing future closed-loop adaptive VR systems that periodically assess changes in user discomfort and enable real-time adaptation throughout immersive VR experiences.





\section{Limitations and Future Work}

The findings of this study should be interpreted in light of several limitations. The relatively small sample size and the use of a single VR environment with a navigation task limit the generalizability of the findings, while the resulting severity group sizes may have reduced the statistical power to detect subtle between-group differences. Moreover, eye-movement events were identified from head-relative gaze without explicitly accounting for head motion; consequently, coordinated eye--head movements may have influenced the observed eye-movement characteristics. Finally, representing early eye-movement behavior using linear trajectories provides an interpretable summary of temporal change but may not capture more complex nonlinear dynamics. Future work should validate these findings across diverse VR applications, jointly analyze eye and head movements, and investigate more flexible temporal modeling approaches.

\section{Conclusion}

This study investigated whether early eye-movement dynamics can characterize the subsequent progression of cybersickness during VR exposure. Unlike previous work that focused primarily on raw eye tracking measures or overall cybersickness severity, this work examined how interpretable gaze trajectories during the first minutes of immersion relate to both discomfort progression and post-completion cybersickness. The results demonstrated that early fixation dispersion trajectories were consistently associated with subsequent discomfort progression, with between-participant differences emerging after the first two minutes of exposure and within-participant differences becoming apparent after approximately four minutes. In contrast, no significant differences were observed between low and high cybersickness severity groups. Together, these findings suggest that early fixation dispersion is more strongly associated with the temporal progression of cybersickness rather than with overall severity. Collectively, these findings highlight the potential of early fixation dispersion as an objective behavioral marker for characterizing cybersickness progression and provide a foundation for future adaptive VR systems capable of monitoring user state and mitigating discomfort before symptoms become severe.


\bibliographystyle{abbrv-doi}

\bibliography{template, references_zotero}
\end{document}